%% file: Interface.tex
\documentclass[modern]{aastex63}
\newcommand{\kms}{\ensuremath{\,\rm{km\,s^{-1}}}}
\input{hidden_commands} % allows the use of \lt and \gt as substitutes to satisfy Endnote
\shorttitle{Search for an Interface}
\shortauthors{Jenkins \& Gry}
\usepackage[utf8]{inputenc}
\usepackage{epsfig}
\begin{document}
\title{In Search of an Interface between Warm and Hot Gas within the Local 
Bubble\footnote{Based on observations with the NASA/ESA Hubble Space Telescope obtained 
from the Data Archive at the Space Telescope Science Institute, which is operated by the 
Associations of Universities for Research in Astronomy, Incorporated, under NASA contract 
NAS5-26555.~ \copyright 2020. The American Astronomical Society. All rights reserved.}}
\author[0000-0003-1892-4423]{Edward B. Jenkins}
\affiliation{Princeton University Observatory,\\ Princeton, NJ 08544}
\author[0000-0003-0788-7452]{Cécile Gry}
\affiliation{Aix Marseille Univ., CNRS, CNES, LAM,\\ Marseille, France}
\correspondingauthor{E. B. Jenkins}
\email{ebj@astro.princeton.edu}
\email{cecile.gry@lam.fr}
\begin{abstract}
We have examined UV spectra recorded by the Space Telescope Imaging Spectrograph (STIS) on 
the {\it Hubble Space Telescope\/} for three stars, HD32309, 41 Ari, and $\eta$~Tel,  that are 
located well inside the boundary of the Local Hot Bubble in our search for absorption features 
of \ion{Si}{4}, \ion{C}{4}, and \ion{N}{5} that could reveal the presence of an interface between 
the local warm ($T\sim 7000$\,K) neutral medium and a more distant hot ($T\sim 10^6$\,K) 
interstellar medium.  In all cases, we failed to detect such ions.  Our most meaningful upper 
limit is that for $\log N$(\ion{C}{4})$\lt 11.86$ toward HD32309, which is below the expectation 
for a sight line that penetrates either a conductive/evaporative interface or a turbulent mixing 
layer.  We offer conjectures on the reasons for these negative results in terms of either a 
suppression of a conductive layer caused by the shielding of the local cloud by other clouds, 
which may make it more difficult for us to sense discrete absorption features from gases at 
intermediate temperatures, or by the presence of a tangential magnetic field at most locations 
on the surface of the local cloud.
\end{abstract}


\section{Background}\label{sec: bkgnd}
The interstellar medium in the plane of our Galaxy occupies an extremely broad range of 
density and temperature. Compensating for these changes in density\footnote{To put these 
contrasts in the perspective of a familiar terrestrial example, the mass density range exceeds 
that spanned by air and lead.}, the accompanying temperatures differ over many orders of 
magnitude, in an inverse fashion.  It is conventional to think of interstellar gaseous matter being 
distributed in diverse, but well defined phases that consist of (1) the hot, collisionally ionized 
($T\sim 10^{6-7}$ K) interstellar medium (HIM) that emits soft X-rays, (2) the warm neutral or 
ionized medium (WNM+WIM) ($6000 \lt T \lt 10^4$ K), (3) the cold ($T\sim 10-200$ K) neutral 
medium (CNM) and (4) very dense molecular star-forming clouds that occupy only a few 
percent of the volume of space.  In some instances the boundaries between these different 
phases can be quite abrupt, and it is not yet clear how they trade matter and entropy. To learn 
more about such interfaces, we can attempt to use UV spectroscopy to investigate individual 
boundaries between the WNM and HIM, so that we can understand better their roles in a 
global environment by enhancing the cooling of the very hot material through either mass 
exchange or conductive and radiative losses (Begelman \& McKee 1990 ; McKee \& Begelman 
1990). This information could help us to construct more accurate accounts on how rapidly cold 
gas clouds within hot gas volumes created by supernova explosions may dissipate and modify 
the evolution of the remnants (Slavin et al. 2017), which in turn might influence the 
morphologies of galaxies and some key aspects of the overall cycle of matter and thermal 
energy within them. 

Many theorists have confronted this issue and have concluded that two basic categories of 
interactions can take place: (1) the establishment of a conductive interface where evaporation 
or condensation can occur (Cowie \& McKee 1977 ; McKee \& Cowie 1977 ; Balbus 1985, 1986 ; 
Ballet et al. 1986 ; Slavin 1989 ; Borkowski et al. 1990 ; Dalton \& Balbus 1993), and (2) a 
turbulent mixing layer (TML), where, as the name implies, the existence of any shear in velocity 
between the phases creates instabilities and mechanically induced chaotic interactions that mix 
the two phases (Begelman \& Fabian 1990 ; Slavin et al. 1993 ; Kwak \& Shelton 2010 ; Fielding 
et al. 2020).  When the velocity differences are extreme, such as with High Velocity Clouds 
(HVCs) entering the Galaxy from the halo, ablation can occur as the clouds pass through a hot 
medium (Kwak et al. 2011 ; Henley et al. 2012).

Observers have attempted to identify these processes chiefly by analyzing interstellar 
absorption features of ions that are most abundant at intermediate temperatures, such as 
\ion{Si}{4}, \ion{C}{4}, \ion{N}{5}, and \ion{O}{6} (Jenkins 1978b, 1978a ; Savage \& Massa 1987 
; Spitzer 1996 ; Sembach et al. 1997 ; Fox et al. 2003 ; Knauth et al. 2003 ; Zsargó et al. 2003 ; 
Indebetouw \& Shull 2004a ; Indebetouw \& Shull 2004b ; Bowen et al. 2008 ; Lehner et al. 
2011 ; Wakker et al. 2012). In some of these studies, comparisons were made with theoretical 
predictions for the column densities across the interfaces.  One shortcoming of these 
investigations is that they have mostly been conducted over very long sight lines, where 
multiple interfaces of different types may be sensed and be mixed with contributions from 
large, contiguous volumes containing radiatively cooling gases or interstellar shocks.  
\newpage
\section{Examining an Interface}\label{sec: exam}

The quest to characterize through observations the nature of the physical processes at a nearby 
boundary of the HIM has proved to be elusive.  Our goal is to observe along a sight line that 
penetrates just one such interface and then examine UV absorption features of atoms in the 
expected intermediate stages of ionization. So far observers have had very limited success in 
this venture. 

The ideal venue for research on a boundary is the local interstellar medium that surrounds the 
Sun. We are located within a WNM cloud, which is diffuse ($n({\rm H})=0.05 – 0.3{\rm cm}^{-
3}$), warm ($T\sim 7000$\,K) and partly ionized ($n(e)/n({\rm H})\sim 0.5$) (Redfield \& Falcon 
2008). It extends out to  distances that range from 9 to 20~pc (Frisch et al. 2011 ; Gry \& Jenkins 
2014).  This cloud in turn is surrounded by a cavity that contains a hot medium (HIM) that is 
collisionally ionized and emits soft X-rays with emission line ratios consistent with a 
temperature $T=1.1\times 10^6\,$K (Sanders et al. 2001 ; Wulf et al. 2019). This cavity, called 
the Local Bubble, extends out to a distance of order 100~pc from us  (McCammon \& Sanders 
1990 ; Snowden et al. 1997 ; Snowden et al. 1998 ; Vergely et al. 2010 ; Welsh et al. 2010b ; 
Lallement et al. 2013 ; Galeazzi et al. 2014 ; Snowden et al. 2014 ; Uprety et al. 2016 ; Liu et al. 
2017).  Discussions in the literature point to an age of the Local Bubble somewhere around 10 
to 20~Myr when it was established and maintained by multiple core-collapse supernova 
explosions that occurred in our general vicinity
(Maíz-Apellániz 2001 ; Benítez et al. 2002 ; Berghöfer \& Breitschwerdt 2002).  One such 
supernova at many tens of pc away from us probably rejuvenated the hot gas as recently as 
$\sim$2~Myr ago, a proposition that is supported by evidence from an excess of $^{60}$Fe 
within a layer in an ocean sediment (Knie et al. 2004), together with its existence in surface 
material on the Moon (Fimiani et al. 2016) and ice cores recovered in Antarctica (Koll et al. 
2019).  An excess of cosmic ray antiprotons and positrons at $E\gtrsim 20\,$GeV offer further 
support for this recent event (Kachelrieß et al. 2015).

We propose that the observation of any sight line that extends from our location out to a 
distance somewhere between about 20 and 100 pc away should in principle enable us to 
achieve our goal of penetrating a single interface (or perhaps a few interfaces). The 
investigations of absorption features from highly ionized atoms that we will present in this 
paper were intended to achieve this goal.

However, there are pitfalls. Main sequence or giant stars with temperatures exceeding 
25,000\,K can produce local \ion{H}{2} regions that have measurable amounts of \ion{Si}{4} and 
\ion{C}{4} that could interfere with attempts to identify those species in an interface (Cowie et 
al. 1981).  For instance, this puts to question the usefulness of the result for $\alpha$~Vir 
(B1\,V, d~=~76~pc) (Huang, et al. 1995).   A way to overcome this problem is to identify an 
absorption feature whose radial velocity is coincident with a known component of the local gas 
and that disagrees with the expected velocity of the \ion{H}{2} region associated with the star.  
Welsh \& Lallement (2005) used the medium resolution echelle mode of the Space Telescope 
Imaging Spectrograph (STIS) on the {\it Hubble Space Telescope\/} ({\it HST}) to detect high ion 
absorptions toward  two out of four early-type stars located beyond the perimeter of the Local 
Bubble, but the line widths were too narrow compared  to the expectation for a local warm-hot 
gas interface.  Detections of \ion{C}{4} features in the spectra of the stars HD158427 (B2\,Vne, 
d~=~74~pc) (Welsh et al. 2010a)\footnote{The robustness of this finding has been questioned 
in Appendix~A of an article by Freire Ferrero et al. (2012).  In addition to pointing out 
complications arising from the star’s immediate environment, these authors suggest that the 
star is somewhat beyond the edge of the Local Bubble.  Unfortunately, this star is too bright for 
a reliable parallax measurement by Gaia.}, $\beta$~CMa (B1\,II-III, d~=~150 pc) (Dupin \& Gry 
1998), and $\epsilon$~CMa (B1.5\,II, d~=~125 pc) (Gry et al. 1995 ; Gry \& Jenkins 2001) 
indicated that the \ion{C}{4} absorptions had line widths consistent with that expected for an 
interface and were detected at a radial velocity that agreed with that of the local warm 
medium.  However, accompanying features of \ion{Si}{4} or \ion{N}{5} were not detected.  In 
other cases, only upper limits for \ion{C}{4} could be obtained (Bertin et al. 1995 ; Holberg et al. 
1999).

White dwarf (WD) stars were once thought to have featureless spectra that could cleanly show 
interstellar lines, but subsequent investigations reveal unexpectedly high metal abundances in 
their atmospheres caused by radiative levitation (Barstow et al. 2003) and pollution by infall or 
outflows of circumstellar matter (Bannister et al. 2003 ; Rafikov \& Garmilla 2012 ; Barstow et 
al. 2014). These lines can seriously compromise attempts to discern interstellar features 
(Lallement et al. 2011), but Barstow et al. (2010) found that in some cases interstellar 
\ion{O}{6} features can be separated from photospheric or circumstellar contributions.  For 
such cases, they occasionally found evidence for \ion{O}{6} absorption arising from inside the 
Local Bubble, but a majority of positive detections had sight lines that extended beyond the 
edge of the Local Bubble.

Freire Ferrero et al. (2012) performed a search for \ion{C}{4} and \ion{Si}{4} lines in a large 
collection of spectra obtained by the {\it International Ultraviolet Explorer\/} ({\it IUE}) with the 
goal of detecting these ions in the local vicinity. They concluded that ideal targets have spectral 
types from B6 to early A and whose spectra do not show peculiarities.  However they only 
detected high-ion lines beyond 90\,pc, i.e. outside the Local Bubble, because the expected 
column density of a single interface is below the detection threshold of IUE spectra.  A 
dedicated high-resolution and high-signal-to-noise spectrum of a nearby, main sequence B6 to 
early A star thus appears necessary to detect a single interface.

A theoretical investigation by Balbus (1986) indicated that a magnetic field that is 
approximately parallel to the surface of an interface should suppress the conduction of heat by 
electrons across the boundary and significantly reduce the column densities of ions at 
intermediate temperatures.  Time dependent cooling calculations by Slavin (1989) and 
Borkowski et al. (1990) revealed how much reduction of \ion{Si}{3}, \ion{Si}{4}, \ion{C}{4}, and 
\ion{N}{5} should take place for different field orientations.  It is possible that this suppression 
effect may explain why a high quality STIS echelle spectrum of $\alpha$~Leo (B8\,Vn, d~=~24 
pc) analyzed by Gry \& Jenkins (2017) failed to show any detectable absorptions from these 
species.  In particular, their conservative upper limit $N$(\ion{C}{4})$\lt 2\times 10^{12}{\rm  
cm}^{-2}$ toward this star is below the predicted value for a conductive interface that has no 
magnetic suppression and is older than a few thousand years.  Gry \& Jenkins (2017) performed 
a detailed analysis of the neutral and partially ionized gas toward $\alpha$~Leo and found that 
it has a filling factor of only 0.13; the remaining span along the sight line must therefore contain 
hot gas, which is supported by the measurements of soft X-ray emission in the neighboring line 
of sight against the nearby, fully opaque Local Leo Cold Cloud, which is situated at a distance of 
$11-24$ pc from us (Snowden et al. 2015).  If turbulent mixing is not occurring and a conductive 
interface is relevant, the failure to detect one toward $\alpha$~Leo may be caused by magnetic 
suppression, since the star is located in a direction that is $81\fdg7$ from the average magnetic 
field that points toward the Galactic coordinates $\ell=36\fdg2$,  $b=49\fdg0$, as indicated by 
polarization measurements of stars within about 40~pc of the Sun (Frisch et al. 2015).  This field 
direction is only $6\fdg7$ from the $3\mu$G field direction very close to the Sun, which is 
evident from the locations in the sky of a ribbon of energetic neutral atoms mapped by the {\it 
Interstellar Boundary Explorer\/} (IBEX) (Zirnstein et al. 2016).

\section{Current Observations}\label{sec: obs}

As with many of the previous studies outlined above, we used the STIS echelle spectrograph, 
and we concentrated on the spectral regions covering the strongest members of the \ion{Si}{4}, 
\ion{C}{4}, and \ion{N}{5} doublets.  Our optimal choice for a target star to search for an 
interface in the foreground was HD32309.  As indicated in Table~\ref{tbl: stars}, it is cool 
enough to ensure that high ions are unlikely to arise from the stellar radiation, it is located in a 
direction that is more likely to be parallel to the local magnetic field, and it is at a distance that 
should extend beyond the edge of the WNM (Wood et al. 2005 ; Gry \& Jenkins 2014) and yet 
does not reach the edge of the Local Bubble.  The star is bright enough in the UV to yield a good 
signal-to-noise ratio (S/N) in a reasonable observing time, yet it is not so bright that it requires a 
neutral density filter to limit the detector count rate to a safe level.  On 2018 February 15 we 
observed this star for 10286\,s during our {\it HST\/} observing program number 15203, where 
we obtained a spectrum covering 1142 – 1687\,\AA\ that ranged from S/N~=45 per resolution 
element at the \ion{N}{5} doublet to 80 at the \ion{C}{4} doublet.  This spectrum was recorded 
using the E140H mode of the STIS echelle spectrograph, which yielded a resolution 
$\lambda/\Delta\lambda=114,000$.  We constructed our observing program so that two 
different central wavelength settings covered each high ion feature, so that we could reduce 
the effects of the detector sensitivity variations that were not fully neutralized by the flat field 
corrections, as described in Section~7.5.3 of the STIS Instrument Handbook (Riley et al. 2019). 

\begin{figure}
\epsscale{1.4}
\plotone{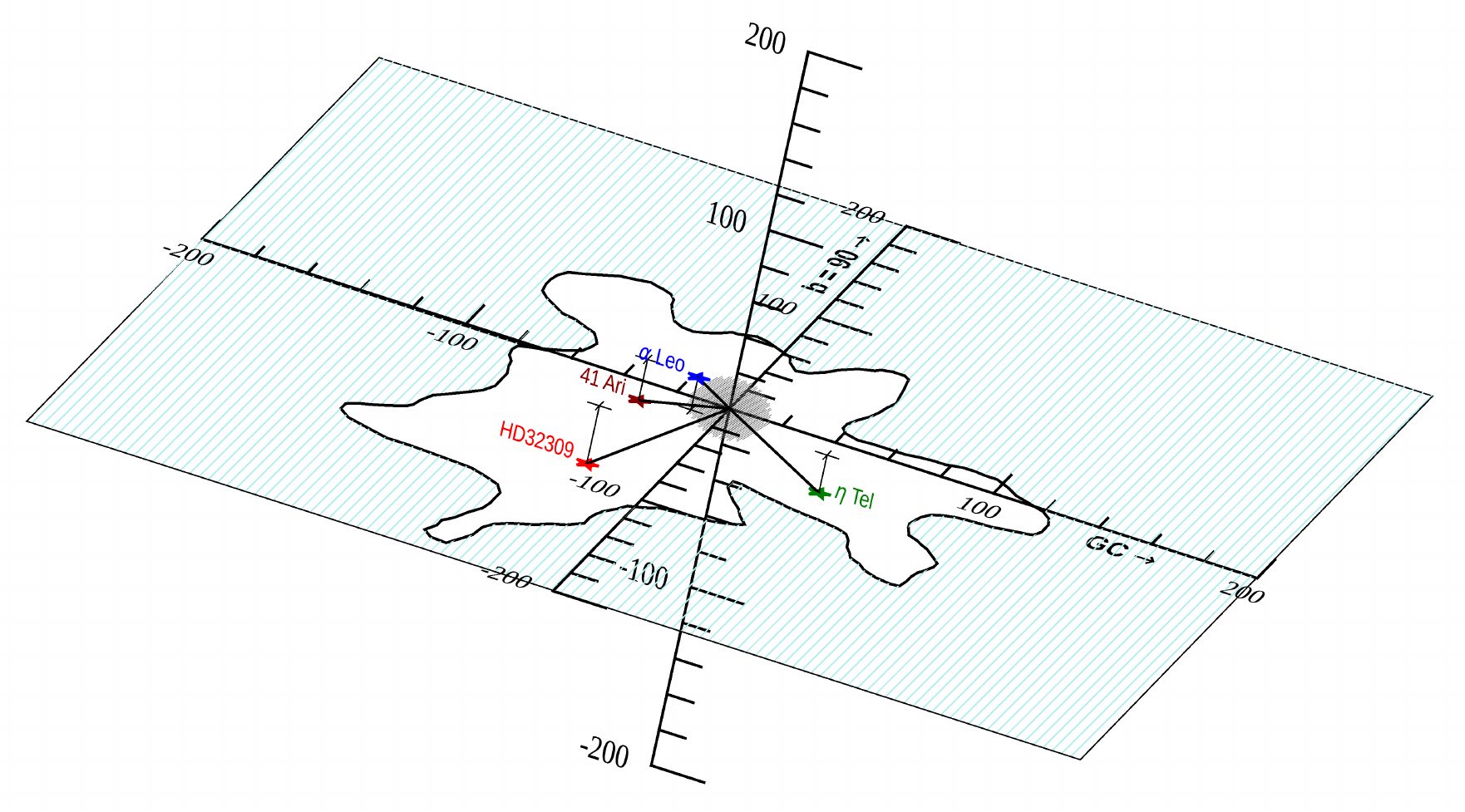}
\caption{A depiction of the three dimensional layout of the sightlines to the three stars under 
investigation, along with that toward $\alpha$~Leo.  The Sun is located at the origin of the axes, 
and a schematic representation of the surrounding WNM cloud complex is illustrated by a tight, 
spherically bounded grouping of thin lines that have an orientation that coincides with the local 
magnetic field direction specified by Frisch et al. (2015).  The Galactic plane is shown with a 
cutout that maps where the inner boundary of the Local Bubble intersects it, as defined by the 
$n$(\ion{Na}{1})$=10^{-9.5}\,{\rm cm}^{-3}$ contour outlined by Welsh et al. (2010b) (see their 
Fig.~12).  The projections of the star locations onto the Galactic plane are shown by thin lines 
and plus signs.  The direction toward the Galactic Center is toward the lower right, and a 
Galactic longitude $b=90\arcdeg$ is toward the upper right.  Numbers on the axes are distances 
in pc.\label{fig: layout}}
\end{figure}

We supplement our observations with data from two other stars that we could download from 
{\it HST\/} public archive: one is a mix of E140M and E140H spectra of the star 41~Ari observed 
for the Advanced Spectral Library~II (ASTRAL): Hot Stars (program 13346, T.~Ayres, PI) 
(Carpenter et al. 2015), and another is $\eta$~Tel, which was observed in the E140H mode for a 
program to search for circumstellar material (program 14207, A.~Roberge, PI).\footnote{ 
Specific details on the dates of the observations, exposure times, spectrograph modes, and 
central wavelength settings that were used for all three stars in this paper can be accessed via 
the following doi collections in the {\it Mikulsky Archive for Space Telescopes\/} ({\it MAST\/}): 
\dataset[10.17909/T9P016]{\doi{10.17909/T9P016}} for the ASTRAL HLSP that holds the 
spectrum of 41~Ari and 
\dataset[10.17909/t9-c4zb-3j36]{\doi{10.17909/t9-c4zb-3j36}} for the individual spectra of 
HD32309 and $\eta$~Tel.  The data were subjected to the standard processing that created {\tt 
.x1d} files in {\it MAST} for HD32309 and $\eta$~Tel and specialized processing (Ayres 2010) for 
overlapping E140M and E140H data for 41~Ari that went into the creation of the HLSP files.}   
These other two stars do not seem as well suited for detecting an interface because their sight 
lines are more likely to penetrate surfaces that have normals that intersect magnetic field lines 
at some appreciable angle.  However, we cannot be certain that any such normals are well 
aligned with the directions to the stars that are being observed, so we concluded that it would 
be beneficial to include them in our study.  Table~\ref{tbl: stars} lists various properties of the 
three stars, and Figure~\ref{fig: layout} shows how the sight lines are oriented relative to the 
average local magnetic field and the inner boundary of the Local Bubble.

\begin{deluxetable}{
r	% HD
c	% name
c	% spectral type
c	% V mag
c	%v_0
c	% l
C	% b
c	% B angle
c	% d
}[h]
\tablewidth{0pt}
\tablecolumns{9}
\tablecaption{Stellar Properties\label{tbl: stars}}
\tablehead{
\colhead{} & \colhead{} & \colhead{Spectral} & \colhead{$V$} 
&\colhead{$v_0$\tablenotemark{a}} & \multicolumn{2}{c}{Galactic Coordinates}  & 
\colhead{$\theta_{\bf B}$\tablenotemark{b}} & \colhead{Distance}\\
\colhead{HD} & \colhead{Name} & \colhead{Type} & \colhead{(mag.)} &\colhead{(\kms)} & 
\colhead{$\ell$} & \colhead{$b$} & \colhead{(deg.)} & \colhead{(pc)}
}
\startdata
\object{HD32309}&\nodata&B9V&4.89&\phm{$-$}19.5&220.3&-32.8&15.5&60\\
\object{HD17573}&41 Ari&B8Vn&3.59&\phm{$-$}17.5&153.0&-28.6&51.6&48\\
\object{HD181296}&$\eta$ Tel&A0V+M7/8V&5.02&$-22.0$&342.9&-26.2&88.9&48\\
\enddata
\tablenotetext{a}{Heliocentric reference velocity established by singly ionized atoms.}
\tablenotetext{b}{Angle (in degrees) between the sight line to the star and the average local 
magnetic field direction within 40 pc of the Sun, as defined by Frisch et al. (2015).}
\end{deluxetable}

\section{Outcomes}\label{sec: outcomes}

For each of the three ions, \ion{Si}{4}, \ion{C}{4}, and \ion{N}{5}, we measured only the 
strongest member of the doublet.  Supplementing such a measurement with a properly 
weighted outcome with the weaker line would provide only a marginal increase in the quality of 
the result.  Absorption features from singly ionized atoms arise from the WNM (or WIM), and 
they provide a zero-point reference $v_0$ for the radial velocity of gas at the base of any 
conduction front or turbulent mixing layer.  Our integration interval for the $W_\lambda$
measurements allowed for a line broadening that combined the effects caused by both thermal 
(at $T=2\times 10^5$\,K) and bulk motion broadenings.  For \ion{Si}{4} we adopted limits equal 
to $-23$ to +28\kms\ relative to $v_0$, but for \ion{C}{4} and \ion{N}{5} this interval was 
increased to $-30$ to +35\kms.  Values of $v_0$ are listed in Table~\ref{tbl: stars}.

In no case did we find what we consider to be a secure, positive detection of a feature.  
Table~\ref{tbl: outcomes} shows the results of our equivalent width measurements.  While 
some measurements indicated positive outcomes at the $\sim 1\sigma$ level, we note that of 
the 8 results shown in the table, only 4 of them indicate positive $W_\lambda$ values.  Thus, 
we have confidence only in stating upper limits for the column densities.  To derive these upper 
limits at the $2\sigma$ level of confidence for $\log(N)$, we applied the relationship for a 
weak, unsaturated line $\log(N)=\log(W_\lambda/\lambda)-\log(f\lambda)+20.053$ to a 
Bayesian analysis of the measured equivalent width $W_\lambda$ and its uncertainty in a 
manner that disallows any negative or unreasonably small outcomes that could arise by chance 
(Bowen et al. 2008, Appendix D).

\begin{deluxetable}{
c	% lambda
c	% S/N
c	% $\log(f/\lambda)$
c	% ion
c	% W/lambda + error
c	% upper limit
}
\tablewidth{0pt}
\tablecolumns{6}
\tablecaption{Column Density Upper Limits\label{tbl: outcomes}}
\tablehead{
\colhead{$\lambda$} & \colhead{S/N\tablenotemark{a}} & \colhead{$\log(f\lambda)$} & 
\colhead{Ion} & \colhead{$W_\lambda$\tablenotemark{b}} & \colhead{$\log(N)$ upper}\\
\colhead{(\AA)} & \colhead{} & \colhead{} & \colhead{} & \colhead{(m\AA)} & \colhead{limit 
($2\sigma$)}
}
\startdata
\cutinhead{HD32309}
1393.755&60&$  2.854 $& \ion{Si}{4} &$   2.4\pm  1.4 $&11.77\\
1548.195&80&$  2.468 $& \ion{C}{4} &$  -0.1\pm  1.3 $&11.86\\
1238.821&45&$  2.286 $&  \ion{N}{5} &$  -2.2\pm  1.5 $&12.02\\
\cutinhead{41 Ari}
1393.755&130&$  2.854 $& \ion{Si}{4} &$   -3.4\pm  1.3$\tablenotemark{c}&11.21\\
1548.195&90&$  2.468 $& \ion{C}{4} &$  1.7\pm  2.7 $&12.27\\
1238.821&140&$  2.286 $&  \ion{N}{5} &$  3.4\pm  3.0 $&12.65\\
\cutinhead{$\eta$ Tel\tablenotemark{d}}
1393.755&40&$  2.854 $& \ion{Si}{4} &$   -1.4\pm  2.7 $&11.78\\
1548.195&60&$  2.468 $& \ion{C}{4} &$  3.9\pm  5.6 $&12.59\\
\enddata
\tablenotetext{a}{Signal-to-noise ratio per detector pixel based on deviations arising from 
photon-counting noise.}
\tablenotetext{b}{Uncertainties are expressed in terms of $1\sigma$ deviations caused by both 
photon-counting noise and an inaccurate continuum placement.}
\tablenotetext{c}{This $-2.6\sigma$ outcome probably arose from the effect of an 
unrecognized detector sensitivity change in a positive direction or hot pixels.}
\tablenotetext{d}{At the location of the \ion{N}{5} doublet, there was an insufficient flux to 
make a meaningful measurement.}
\end{deluxetable}
\newpage
\section{Implications}\label{sec: implications}
\subsection{Conduction Front}\label{sec: conduction front}

We start by considering the static case where a conduction/evaporation front could be 
established between the HIM and WNM (or WIM).  The theoretical studies of magnetized 
conductive interfaces by Slavin (1989) and Borkowski et al. (1990) showed that of the three ions 
that we measured, \ion{C}{4} is expected to exhibit the largest column density.  Our $2\sigma$ 
upper limit  $N$(\ion{C}{4})$\lt 7\times 10^{11}{\rm cm}^{-2}$ for this ion in front of HD32309 
is well below a value $N$(\ion{C}{4})$=3-4\times 10^{12}{\rm cm}^{-2}$ that they computed for 
the case where the electron conduction perpendicular to the front is not inhibited by a 
magnetic field.   Borkowski et al. (1990) indicated that this condition can be reached for a front 
whose age $t\gt  10^4$\,yr.  As we indicated earlier, the HIM within the Local Bubble is much 
older than this time threshold.   Our upper limits for $N$(\ion{C}{4}) toward the other two stars 
are less constraining than that toward HD32309.  For \ion{N}{5}, the models indicate a column 
density equal to $1\times 10^{12}{\rm cm}^{-2}$ after $10^5{\rm yr}$, which is close to our 
upper limit for $N$(\ion{N}{5}) toward HD32309 but well below that for 41~Ari. Our upper 
limits for $N$(\ion{Si}{4}) for all three stars are not a real constraint in the current context, but 
nevertheless they are consistent with our apparent nondetections of \ion{C}{4} and \ion{N}{5}.

As we indicated in Section~\ref{sec: exam}, a major factor that could change the conclusions 
about column densities is the direction of a magnetic field relative to the surface normal of the 
front.  The calculations mentioned earlier indicated that if this angle is equal to or exceeds 
60\arcdeg,  $N$(\ion{C}{4}) drops below $1\times 10^{12}{\rm cm}^{-2}$ (Borkowski et al. 
1990) or $1.5\times 10^{12}{\rm cm}^{-2}$ (Slavin 1989).  Our choice for observing HD32309 
was based on the supposition that a more favorable orientation of the field could improve our 
chance of detecting a highly ionized atom within the front.  Table~\ref{tbl: stars} shows that the 
direction toward this star is only $\theta_{\bf B}=15\fdg5$ from a representative local magnetic 
field direction determined by Frisch et al. (2015).  Supporting the proposition that this angle is 
probably small are the highly precise measurements of polarization carried out by Piirola et al. 
(2020) that indicated stars very close to the direction of HD32309 showed no detectable 
polarization arising from a transverse alignment of foreground dust grains (see their Fig.~5).  
However, as we stated earlier, we have no way of knowing that the surface normal of a front is 
nearly aligned with the direction of our sight line.

\subsection{Turbulent Mixing Layer (TML)}\label{sec: TML}

If the interface between the cool and hot gas is subject to turbulence, either intrinsic to the hot 
gas or generated by a velocity shear that creates a Kelvin-Helmholtz instability at the surface, a 
TML may be the dominant structure (Begelman \& Fabian 1990 ; Slavin et al. 1993).  As with the 
conduction front discussed in the previous section, there should exist gas at intermediate 
temperatures that could hold the three ions that we have investigated.  Kwak \& Shelton (2010) 
have created hydrodynamic simulations that included radiative cooling and
non-equilibrium ionization equilibria to compute how the ion column densities through such 
layers build up with time.  As with the conduction front, \ion{C}{4} should exhibit a column 
density greater than either \ion{Si}{4} or \ion{N}{5}.  Their model with a 50\kms\ shear indicates 
that $N$(\ion{C}{4}) could reach as high as $9\times 10^{12}{\rm cm}^{-2}$, but only after 
about 35~Myr.  A buildup to $10^{12}{\rm cm}^{-2}$ could occur within 10~Myr.  This column 
density could be achieved in only 4~Myr if the velocity shear were as high as 100\kms.  
However, we point out that the calculations of Kwak \& Shelton (2010) applied to a thermal 
pressure of the two media \lt 10\% of that within the Local Bubble.  Simulations of TMLs by Ji et 
al. (2019) that applied to even lower pressures indicated that the ion column densities scale 
weakly with pressure, so the results of Kwak \& Shelton (2010) probably underestimate the 
column densities that we should anticipate for a TML inside the Local Bubble.

\section{Possible Trivial Reasons for a Non-detection of \ion{C}{4} toward HD32309}\label{sec: 
trivial_reasons}
\subsection{Is the Foreground Completely Filled with a Partly Ionized WNM?}\label{sec: 
foreground}

Our \ion{C}{4} upper limit determination for HD32309 is our most compelling non-detection of 
an interface.  One lingering doubt we might have is that the sight line to this star might be 
completely filled with a low density medium that has a temperature of order $10^4$\,K or less, 
which would indicate that any interface with the HIM is beyond the star.  To investigate this 
issue, we need to have a better understanding about the gas in front of this star.  
Unfortunately, we are unable to directly measure $N$(\ion{H}{1}). The star has an effective 
temperature low enough to create a stellar Ly$\alpha$ feature that is so large that we are 
unable to measure the interstellar contribution.   For this reason, plus a need to determine the 
amount of ionized gas that may be present, we must use an indirect argument to help us to 
understand the probable nature of the material in front of HD32309.

Using f-values from Kisielius et al. (2014), we have determined a column density 
$N$(\ion{S}{2})$=1.46\times 10^{13}{\rm cm}^{-2}$ from the 1253.8 and 1259.5\,\AA\ features 
in the spectrum of HD32309.  For the medium in front of $\alpha$~Leo (Galactic coordinates 
$\ell=226\fdg4$, $b=+48\fdg9$, distance $d=24$\,pc), situated at $81\fdg9$ from HD32309,  
Gry \& Jenkins (2017) determined\footnote{We have increased their column density by 5.5\% 
to account for the difference between the f-values listed by Morton (2003)  compared to those 
determined by Kisielius et al. (2014).} that $N$(\ion{S}{2})$=2.53\times 10^{13}{\rm cm}^{-2}$, 
i.e., a factor of 1.7 greater than the column density that we find for HD32309.  If we assume 
that the S abundances and the ionization fractions of S and H for the gas in front of 
$\alpha$~Leo and HD32309 are about the same because both sight lines are dominated by 
absorption in the local cloud as defined by Gry \& Jenkins (2014), we can take the column 
density $N$(\ion{H}{1}~+~\ion{H}{2})$=2.8\times 10^{18}{\rm cm}^{-2}$ toward $\alpha$~Leo 
and divide it by 1.7 to obtain an estimate $1.65\times 10^{18}{\rm cm}^{-2}$ for the total 
amount of hydrogen (neutral and ionized) in front of HD32309.  If instead of being mostly 
confined within the local cloud, this gas were distributed uniformly along the 60~pc long sight 
line toward HD32309, the  volume density would equal 
$n$(\ion{H}{1}~+~\ion{H}{2})$=8.9\times 10^{-3}{\rm cm}^{-3}$.  When we add electrons and 
helium atoms, the total particle density would reach 0.013$~{\rm cm}^{-3}$.  For a temperature 
$T=6000$\,K (again for the gas in front of $\alpha$~Leo), the thermal pressure $nkT=1.1\times 
10^{-14}{\rm dyne~cm}^{-2}$.  Without other means of pressure support, material at this 
pressure could not survive in the hot gas environment with a thermal pressure $1.4\times 10^{-
12}{\rm dyne~cm}^{-2}$ (Snowden et al. 2014).  Thus, we propose that it is unlikely that the 
entire sight line toward HD32309 is completely filled with a WNM.
\newpage
\subsection{Could Much of the Carbon be Depleted onto Dust Grains?}\label{sec: depletion}

The lack of an observed amount of \ion{C}{4} at or above the expectation for a front in the 
foreground of HD32309 might arise from a gas-phase carbon abundance below the solar value.  
Again, we can rely on evidence elsewhere in the local medium for guidance on this issue.  First, 
the abundance of \ion{C}{2} relative to that of H toward $\alpha$~Leo is 0.06~dex higher than 
the protosolar C/H value\footnote{Lodders (2003) estimated that the effects of gravitational 
settling result in protosolar abundances being higher than solar photospheric abundances by 
0.07~dex.  In some cases, the theoretical models for both conduction fronts and turbulent 
mixing layers may have used photospheric abundances, some of which are outdated.}  (Gry \& 
Jenkins 2017). Second,  Redfield \& Linsky (2004a) reported an extensive study of abundances 
of gases in front of stars within 100~pc.  In our interpretation of their findings, we can use 
\ion{N}{1} as a proxy for \ion{H}{1}, since its abundance is slightly lower ($-$0.11~dex) than the 
protosolar value over a wide range of conditions in the ISM (Jenkins 2009).  There are 9 
sightlines in their investigation that had measurements of column densities for both \ion{C}{2} 
and \ion{N}{1}.  We have determined a weighted mean for the quantity $\log N$(\ion{C}{2})$-
\log N$(\ion{N}{1}) equal to $0.98\pm 0.10$ for these stars, plus one limit \gt 0.60.  Based on a 
protosolar C/N = +0.56~dex (Lodders 2003) and allowing for the small depletion of N in the ISM, 
the observed \ion{C}{2} appears to be 0.31~dex higher than expected.  We recognize that some 
of the \ion{C}{2} may come from some regions that are fully ionized, where the N is also 
ionized.  From these results, we conclude that in the low density local ISM it is unlikely that a 
significant fraction of the carbon atoms are sequestered into dust grains.

\section{Discussion}

Our inability to detect an interface in three different directions, along with our earlier null result 
for $\alpha$~Leo (Gry \& Jenkins 2017), deprives us from being able to investigate properties of 
such a boundary or even to confirm its existence.  As we mentioned in Section~\ref{sec: exam}, 
the only cases where we are aware of possible detections of an interface, indicated by solely by 
the most conspicuous ion \ion{C}{4}, are in the spectra of HD158427 (Welsh et al. 2010a), 
$\beta$~CMa (Dupin \& Gry 1998), and $\epsilon$~CMa (Gry et al. 1995 ; Gry \& Jenkins 2001), 
with no supporting evidence from other highly ionized species.  A study by Barstow et al. (2010) 
of \ion{O}{6} absorptions that could be securely identified as interstellar in origin in the spectra 
of a large number of white dwarf stars revealed very few detections that were clearly inside the 
Local Bubble (the closest case was a star at a distance of 58~pc).  Welsh \& Lallement (2008) 
performed a survey of \ion{O}{6} for 17 B-type stars with distances in the range $20 \lt d \lt 
200$\,pc.  For their sample, they obtained only 5 significant detections with 
$N$(\ion{O}{6})$\gtrsim 10^{13}{\rm cm}^{-2}$, all of which were located at $d\gt 80$\,pc.

The geometrical arrangements of clouds and magnetic fields that could influence our transition 
from a WNM to the HIM may be complex, and such conditions may have a strong influence on 
our ability to detect an interface with highly ionized atoms.  We draw attention to two factors 
that could lead to our nondetections.

Our first consideration is that a conduction front can be suppressed by the presence of other 
nearby clouds beyond the edge of our cloud.  Balbus (1985) pointed out that clouds within an 
ensemble can shield each other from the influence of an external hot medium in much the 
same manner that an arrangement electrical conductors can shield each other in the presence 
of an external electrostatic field.  Studies of UV absorptions by Redfield \& Linsky (2004a, 2008) 
indicated that most of the WNM is within about 15\,pc of the Sun, but a survey of \ion{Ca}{2} 
toward a very large number of stars within the Local Bubble by Welsh et al. (2010b) indicates 
that sparse wisps of warm material exist beyond 15\,pc.  According to Balbus (1985), if many 
clouds with radii $a$ are situated within a spherical region of radius $R$, their volume filling 
factor $f$ must satisfy the condition that $f\ll (2/3)(a/R)^2$ for them not to be influential in 
mutually suppressing their interfaces.  This effect could modify the nature of our transition 
from the WNM to the HIM and make it more difficult to sense the presence of gases at 
intermediate temperatures.

The second consideration is the geometrical configuration of the WNM and its accompanying 
internal magnetic field directions.  We may have been naïve in assuming that the magnetic field 
direction in the WNM is almost the same everywhere.  Slavin (2020) conducted a series of MHD 
simulations of initially cold clouds subjected to multiple supernova shocks (the same shocks 
that created the hot gas in the Local Bubble).  After compression by the shocks, the clouds 
rebounded and were heated to their current temperatures.  During the expansion, if the initial 
ratio of thermal to magnetic pressures $\beta\gg 1$ interior to the cloud, we might expect that 
the cloud’s elevated pressure could compress some portion of the internal magnetic field 
against the perimeter as it pushes against the opposing external pressure of the surrounding 
HIM.  Ultimately, this action might align the field lines such that they are nearly parallel to the 
surface.  If we were to propose that this  tangential field strength is of order $5\mu$G, its 
pressure $B^2/8\pi=1\times 10^{-12}{\rm dyne~cm}^{-2}$, when combined with typical 
thermal plus turbulent pressures of the local cloud complex $nkT=3.3\times 10^{-13}{\rm 
dyne~cm}^{-2}$ (Redfield \& Linsky 2004b) could stabilize the medium against the external 
pressure which is estimated to be $1.4\times 10^{-12}{\rm dyne~cm}^{-2}$ from the 
surrounding HIM (Snowden et al. 2014).  As we indicated earlier, if this compressed tangential 
field is stronger than the radial one, which is likely, the resulting alignment could force the heat 
transfer across the cloud boundary to be inhibited and substantially reduce the presence of 
highly ionized atoms in the front.  This field configuration may explain why we were unable to 
detect a conduction front.

For the case of a turbulent mixing layer (TML), we might question whether or not broadening 
due to the turbulence might make the absorption depths of the lines too small to detect and 
our measurement velocity interval inappropriately too narrow.  Numerical experiments of TMLs 
by Fielding et al. (2020) indicate that the turbulent velocity is only about 0.1 to 0.2 times the 
differential shear velocity.  One possible clue about the motion of the local medium relative to 
the HIM may arise from the interpretation of the cloud’s internal kinematics by Gry \& Jenkins 
(2014), suggesting that the local medium may be undergoing compression by the ram pressure 
coming from the direction $\ell=174\arcdeg$, $b=-12\arcdeg$.  This direction is 47\arcdeg\ 
from the sight line to HD32309, so a transverse flow (of uncertain magnitude) could be present 
and might create a TML.

An important issue that directly applies to the local cloud surrounding the Sun is a surprisingly 
high fraction of helium that is in the singly ionized state, as indicated by EUV observations of 
white dwarfs (Dupuis et al. 1995 ; Barstow et al. 1997 ; Cruddace et al. 2002).  These results are 
difficult to explain in terms of just the EUV radiation from known stellar sources in the local 
region (Vallerga 1998).  Slavin \& Frisch (2002) proposed that, in addition to ionizing radiation 
from stars, the EUV and soft X-ray emissions by a surrounding conductive interface are also 
necessary.  If evidence for an interface in the local area is lacking, there may be a need to
re-examine the explanation for the He ionization. 

\acknowledgments
We thank the referee for a useful and prompt report.  This research was based on observations 
with the NASA/ESA Hubble Space Telescope obtained from the {\it Mikulski Archive for Space 
Telescopes\/} (MAST) maintained at the Space Telescope Science Institute (STScI), which is 
operated by the Association of Universities for Research in Astronomy, Incorporated, under 
NASA contract NAS5-26555.  Support for program number HST-GO-15203.002-A was 
provided through a grant to Princeton University from the STScI under NASA contract NAS5-
26555.
\facilities{HST (STIS)}

\end{document}

%% file: hidden_commands.tex
\newcommand{\lt}{\ensuremath{<}} 
\newcommand{\gt}{\ensuremath{>}}

%% file: Interface.bbl
\begin{references}
\reference{7541} Ayres, T. R. 2010, ApJS, 187, 149
\reference{6595} Balbus, S. A. 1985, ApJ, 291, 518
\reference{8720} ---. 1986, ApJ, 304, 787
\reference{1946} Ballet, J., Arnaud, M., \& Rothenflug, R. 1986, A\&A, 161, 12
\reference{7115} Bannister, N. P., Barstow, M. A., Holberg, J. B., \& Bruhweiler, F. C. 2003, 
MNRAS, 341, 477
\reference{3503} Barstow, M. A., Dobbie, P. D., Holberg, J. B., Hubeny, I., \& Lanz, T. 1997, 
MNRAS, 286, 58
\reference{5154} Barstow, M. A., Good, S. A., Holberg, J. B., et al. 2003, MNRAS, 341, 870
\reference{7240} Barstow, M. A., Boyce, D. D., Welsh, B. Y., et al. 2010, ApJ, 723, 1762
\reference{8132} Barstow, M. A., Barstow, J. K., Casewell, S. L., Holberg, J. B., \& Hubeny, I. 
2014, MNRAS, 440, 1607
\reference{36} Begelman, M. C., \& Fabian, A. C. 1990, MNRAS, 244, 26P
\reference{39} Begelman, M. C., \& McKee, C. F. 1990, ApJ, 358, 375
\reference{8806} Benítez, N., Maíz-Apellániz, J., \& Canelles, M. 2002, PhRvL, 88, 081101 
\reference{4928} Berghöfer, T. W., \& Breitschwerdt, D. 2002, A\&A, 390, 299
\reference{3135} Bertin, P., Vidalmadjar, A., Lallement, R., Ferlet, R., \& Lemoine, M. 1995, 
A\&A, 302, 889
\reference{2186} Borkowski, K. J., Balbus, S. A., \& Fristrom, C. C. 1990, ApJ, 355, 501
\reference{6557} Bowen, D. V., Jenkins, E. B., Tripp, T. M., et al. 2008, ApJS, 176, 59
\reference{8328} Carpenter, K. G., Ayres, T. R., \& ASTRAL~Science~Team. 2015,  18th 
Cambridge Workshop on Cool Stars, Stellar Systems, and the Sun, eds. G. T. van Belle, \& H. C. 
Harris  18.
\reference{1692} Cowie, L. L., \& McKee, C. F. 1977, ApJ, 211, 135
\reference{1152} Cowie, L. L., Taylor, W., \& York, D. G. 1981, ApJ, 248, 528
\reference{4947} Cruddace, R. G., Kowalski, M. P., Yentis, D., et al. 2002, ApJ, 565, L47
\reference{2416} Dalton, W. W., \& Balbus, S. A. 1993, ApJ, 404, 625
\reference{4337} Dupin, O., \& Gry, C. 1998, A\&A, 335, 661
\reference{3171} Dupuis, J., Vennes, S., Bowyer, S., Pradhan, A. K., \& Thejll, P. 1995, ApJ, 455, 
574
\reference{9650} Fielding, D. B., Ostriker, E. C., Bryan, G. L., \& Jermyn, A. S. 2020, ApJ, 894, 
L24
\reference{8802} Fimiani, L., Cook, D. L., Faestermann, T., et al. 2016, PhRvL, 116, 151104
\reference{5075} Fox, A. J., Savage, B. D., Sembach, K. R., et al. 2003, ApJ, 582, 793
\reference{7647} Freire Ferrero, R., Morales Durán, C., Halbwachs, J. L., \& Cabo Cubeiro, A. M. 
2012, AJ, 143, 28
\reference{7487} Frisch, P. C., Redfield, S., \& Slavin, J. D. 2011, ARA\&A, 49, 237
\reference{8691} Frisch, P. C., Berdyugin, A., Piirola, V., et al. 2015, ApJ, 814
\reference{8658} Galeazzi, M., Chiao, M., Collier, M. R., et al. 2014, Natur, 512, 171
\reference{3097} Gry, C., Lemonon, L., Vidal-Madjar, A., Lemoine, M., \& Ferlet, R. 1995, A\&A, 
302, 497
\reference{3905} Gry, C., \& Jenkins, E. B. 2001, A\&A, 367, 617
\reference{8185} Gry, C., \& Jenkins, E. B. 2014, A\&A, 567, A58
\reference{8808} ---. 2017, A\&A, 598, A31
\reference{7696} Henley, D. B., Kwak, K., \& Shelton, R. L. 2012, ApJ, 753, 58
\reference{4574} Holberg, J. B., Bruhweiler, F. C., Barstow, M. A., \& Dobbie, P. D. 1999, ApJ, 
517, 841
\reference{5523} Indebetouw, R., \& Shull, J. M. 2004a, ApJ, 607, 309
\reference{8718} Indebetouw, R., \& Shull, J. M. 2004b, ApJ, 605, 205
\reference{1022} Jenkins, E. B. 1978a, ApJ, 219, 845
\reference{1334} ---. 1978b, ApJ, 220, 107
\reference{6999} ---. 2009, ApJ, 700, 1299
\reference{9338} Ji, S., Oh, S. P., \& Masterson, P. 2019, MNRAS, 487, 737
\reference{9653} Kachelrieß, M., Neronov, A., \& Semikoz, D. V. 2015, PhRvL, 115, 181103
\reference{8007} Kisielius, R., Kulkarni, V. P., Ferland, G. J., Bogdanovich, P., \& Lykins, M. L. 
2014, ApJ, 780, 76
\reference{5208} Knauth, D. C., Howk, J. C., Sembach, K. R., Lauroesch, J. T., \& Meyer, D. M. 
2003, ApJ, 592, 964
\reference{7641} Knie, K., Korschinek, G., Faestermann, T., et al. 2004, PhRvL, 93, 171103
\reference{9652} Koll, D., Korschinek, G., Faestermann, T., et al. 2019, PhRvL, 123, 072701
\reference{7153} Kwak, K., \& Shelton, R. L. 2010, ApJ, 719, 523
\reference{8201} Kwak, K., Henley, D. B., \& Shelton, R. L. 2011, ApJ, 739, 30
\reference{7393} Lallement, R., Welsh, B. Y., Barstow, M. A., \& Casewell, S. L. 2011, A\&A, 
533, A140
\reference{7944} Lallement, R., Vergely, J.-L., Valette, B., et al. 2013, A\&A, 561, A91
\reference{7252} Lehner, N., Zech, W. F., Howk, J. C., \& Savage, B. D. 2011, ApJ, 727, 46
\reference{8865} Liu, W., Chiao, M., Collier, M. R., et al. 2017, ApJ, 834
\reference{5604} Lodders, K. 2003, ApJ, 591, 1220
\reference{3329} Maíz-Apellániz, J. 2001, ApJ, 560, L83
\reference{59} McCammon, D., \& Sanders, W. T. 1990, ARA\&A, 28, 657
\reference{1693} McKee, C. F., \& Cowie, L. L. 1977, ApJ, 215, 213
\reference{38} McKee, C. F., \& Begelman, M. C. 1990, ApJ, 358, 392
\reference{5404} Morton, D. C. 2003, ApJS, 149, 205
\reference{9647} Piirola, V., Berdyugin, A., Frisch, P. C., et al. 2020, A\&A, 635, A46
\reference{8719} Rafikov, R. R., \& Garmilla, J. 2012, ApJ, 760, 123
\reference{5392} Redfield, S., \& Linsky, J. L. 2004a, ApJ, 602, 776
\reference{5603} ---. 2004b, ApJ, 613, 1004
\reference{6549} ---. 2008, ApJ, 673, 283
\reference{6710} Redfield, S., \& Falcon, R. E. 2008, ApJ, 683, 207
\reference{1495} Riley, A., Branton, D., Carlberg, J., et al. 2019, STIS Instrument Handbook 
(19.0 ed.; Baltimore: STScI)
\reference{3990} Sanders, W. T., Edgar, R. J., Kraushaar, W. L., McCammon, D., \& 
Morgenthaler, J. P. 2001, ApJ, 554, 694
\reference{1148} Savage, B. D., \& Massa, D. 1987, ApJ, 314, 380
\reference{4085} Sembach, K. R., Savage, B. D., \& Tripp, T. M. 1997, ApJ, 480, 216
\reference{1773} Slavin, J. D. 1989, ApJ, 346, 718
\reference{2433} Slavin, J. D., Shull, J. M., \& Begelman, M. C. 1993, ApJ, 407, 83
\reference{3974} Slavin, J. D., \& Frisch, P. C. 2002, ApJ, 565, 364
\reference{9655} Slavin, J. D., Smith, R. K., Foster, A., et al. 2017, ApJ, 846, 77
\reference{9656} Slavin, J. D. 2020,  arXiv:  2007.07113
\reference{3387} Snowden, S. L., Egger, R., Freyberg, M. J., et al. 1997, ApJ, 485, 125
\reference{3814} Snowden, S. L., Egger, R., Finkbeiner, D. P., Freyberg, M. J., \& Plucinsky, P. P. 
1998, ApJ, 493, 715
\reference{8242} Snowden, S. L., Chiao, M., Collier, M. R., et al. 2014, ApJ, 791, L14
\reference{8516} Snowden, S. L., Heiles, C., Koutroumpa, D., et al. 2015, ApJ, 806, 119
\reference{344} Spitzer, L. 1996, ApJ, 458, L29
\reference{9643} Uprety, Y., Chiao, M., Collier, M. R., et al. 2016, ApJ, 829, 83
\reference{3481} Vallerga, J. 1998, ApJ, 497, 921
\reference{7141} Vergely, J. L., Valette, B., Lallement, R., \& Raimond, S. 2010, A\&A, 518, A31
\reference{7570} Wakker, B. P., Savage, B. D., Fox, A. J., \& Benjamin, R. 2012, ApJ, 749, 157
\reference{6057} Welsh, B. Y., \& Lallement, R. 2005, A\&A, 436, 615
\reference{6835} ---. 2008, A\&A, 490, 707
\reference{7122} Welsh, B. Y., Wheatley, J., Siegmund, O. H. W., \& Lallement, R. 2010a, ApJ, 
712, L199
\reference{7091} Welsh, B. Y., Lallement, R., Vergely, J. L., \& Raimond, S. 2010b, A\&A, 510, 
A54
\reference{5915} Wood, B. E., Redfield, S., Linsky, J. L., Müller, H. R., \& Zank, G. P. 2005, ApJS, 
159, 118
\reference{9644} Wulf, D., Eckart, M. E., Galeazzi, M., et al. 2019, ApJ, 884, 120
\reference{8716} Zirnstein, E. J., Heerikhuisen, J., Funsten, H. O., et al. 2016, ApJ, 818, L18
\reference{5204} Zsargó, J., Sembach, K. R., Howk, J. C., \& Savage, B. D. 2003, ApJ, 586, 1019
\end{references}
